\documentstyle[amssymb,preprint,aps]{revtex}

\newtheorem{theorem}{Theorem}
\newtheorem{acknowledgement}[theorem]{Acknowledgement}

\begin{document}
\title{Reshaping-induced spatiotemporal chaos in driven, damped sine-Gordon
systems}
\author{R. Chac\'{o}n }
\address{Departamento de Electr\'{o}nica e Ingenier\'{\i}a Electromec\'{a}%
nica,\\
Escuela de Ingenier\'{\i}as Industriales, Universidad de Extremadura, E-06071%
\\
Badajoz, Spain}
\date{\today }
\maketitle

\begin{abstract}
Spatiotemporal chaos arising from the competition between
sine-Gordon-breather and kink-antikink-pair solitons by reshaping an ac
force is demonstrated. After introducing soliton collective coordinates,
Melnikov's method is applied to the resulting effective equation of motion
to estimate the parameter-space regions of the ac force where homoclinic
bifurcations are induced. The analysis reveals that the chaos-order
threshold exhibits sensitivity to small changes in the force shape. Computer
simulations of the sine-Gordon system show good agreement with these
theoretical predictions.
\end{abstract}

\section{\protect\bigskip Introduction}

Although nonlinearity is an ubiquitous and fascinating feature of real-world
dynamical phenomena, some aspects of such a fundamental property remain
poorly explored, partly because they are rarely incorporated into the
mathematical models which aim to describe even the simplest of such
phenomena. In the context of nonautonomous systems, one of these aspects is
the nature of the periodic excitations which can act on a given nonlinear,
dissipative or Hamiltonian, system. Since nonlinear systems represent the
general case, it would be natural to model periodic excitations (external or
parametric) by using periodic functions that are solutions of (physically
relevant) nonlinear equations, instead of harmonic functions which are
solutions of linear systems, even though the latter have been overwhelmingly
used up to now. To consider nonlinear periodic excitations implies enlarging
the amplitude-period parameter space to include the parameters, $\alpha _{i}$%
, that control the excitation shape. Note that, in the general case, these
parameters $\alpha _{i}$ are the (normalized) Fourier coefficients of the
periodic function. In physical terms this means that, for a fixed amplitude
and period, the parameters $\alpha _{i}$ are responsible for the temporal
rate at which energy is transferred from the excitation mechanism to the
system. In the context of systems described by ordinary differential
equations, this idea has led to the demonstration of the existence of new
generic {\it temporal} chaos$\leftrightarrow $order routes by changing only
the shape of a nonlinear periodic excitation [1-3]. 

In this present work, the onset of {\it spatio-temporal} chaos associated
with the competition between sine-Gordon-breathers and kink-antikink-pairs
is investigated in a one-dimensional sine-Gordon (sG) system 
\begin{equation}
\Phi _{tt}-\Phi _{xx}+\sin \Phi =-\delta \Phi _{t}+\gamma 
\mathop{\rm sn}%
\left[ 4K(m)t/T;m\right] ,  \eqnum{1}
\end{equation}%
when {\it only} the excitation shape is varied. Here, $%
\mathop{\rm sn}%
\left( u;m\right) $ is the Jacobian elliptic function (JEF) of parameter $m$
($K(m)$ is the complete elliptic integral of the first kind) used {\it solely%
} as an example to illustrate the reshaping-induced order-chaos route. When $%
m=0$, then $%
\mathop{\rm sn}%
\left[ 4K(m=0)t/T;m=0\right] =\sin (2\pi t/T)$, i.e., one recovers the
previously studied case of harmonic excitation [4]. This is relevant in
comparing the structural stability of the sG system when solely the
excitation shape is varied. In the other limit, $m=1$, one obtains the
Fourier expansion of the square wave function: $%
\mathop{\rm sn}%
\left[ 4K(m=1)t/T;m=1\right] =\left( 4/\pi \right) \sum_{n=0}^{\infty
}\left( 2n+1\right) ^{-1}\sin \left[ \left( 2n+1\right) 2\pi t/T\right] $.
From an experimental standpoint, the choice of the above JEF is motivated
for the fact that only the excitation shape is varied by increasing $m$ from
0 to 1, and there is thus a smooth transition from a sine function to a
square wave. While these two limiting shapes are readily implemented in
experiments, no theoretical analysis connecting them has been developed up
to now in the context of sG systems.

The organization of the paper is as follows. In Section 2, a collective
coordinate approach [4] is used to obtain an effective equation of motion
for the collective variables. In Section 3, Melnikov's method is applied to
the resulting effective equation of motion to estimate the parameter-space
regions of the elliptic ac force where the onset of chaotic instabilities is
expected. Numerical evidence supporting the theoretical predictions from
previous sections is presented in Section 4. Finally, Section 5 includes a
summary of the findings.

\bigskip 

\bigskip 

\bigskip 

\bigskip 

\section{\protect\bigskip Collective coordinates}

In this present work, a single-breather excitation is assumed to be present
in the absence of dissipation and forcing ($\delta =\gamma =0)$. It is well
known that the simultaneous action of dissipative $\left( \delta >0\right) $
and spatially uniform harmonic forces $\left( \gamma \sin \left( 2\pi
t/T\right) \right) $ can break the breather into a kink-antikink pair, which
can then recombine into a breather soliton [4,5]. For fixed dissipation,
this process may occur repeatedly yielding an intermittent chaotic sequence
for certain values of amplitude and period. The above elliptic ac driving
permits one to study such a competition between the breather and the
kink-antikink pair for parameter values belonging to {\it larger} regions of
the amplitude-period parameter space. To this end, one makes a severe mode
truncation to the breather collective coordinates [6,7], which yields a set
of ordinary differential equations capable of being studied with the aid of
Melnikov's method [8,9]. Such an effective equation exhibits a separatrix
dynamics which is equivalent to the separatrix dynamics of the full
sine-Gordon equation. In particular, homoclinic bifurcations correspond to
(chaotic) intermittent binding and unbinding of the kink-antikink pair. In
order to determine the equations of motion for the breather collective
modes, one first needs the breather and kink-antikink solutions to the
unperturbed sine-Gordon equation ($\delta =\gamma =0$ in Eq. (1)) [4].

Breather: 
\begin{eqnarray}
U_{B}(x,t) &\equiv &4\tan ^{-1}\left\{ \frac{\tan \theta \sin \left[ v\left(
t\right) \right] }{\cosh \left[ 2k\left( x-x_{0}\right) \right] }\right\} , 
\nonumber \\
v(t) &\equiv &v_{0}+t\cos \theta ,  \nonumber \\
k &\equiv &\frac{1}{2}\sin \theta ,  \nonumber \\
E &\equiv &16\sin \theta .  \eqnum{2}
\end{eqnarray}

Kink-antikink pair ($K\stackrel{\_}{K}$): 
\begin{eqnarray}
U_{K\stackrel{\_}{K}}(x,t) &\equiv &4\tan ^{-1}\left\{ \frac{\sinh \left[
u\left( t-t_{0}\right) /(1-u^{2})^{1/2}\right] }{u\cosh \left[ \left(
x-x_{0}\right) /(1-u^{2})^{1/2}\right] }\right\} ,  \nonumber \\
E_{K\stackrel{\_}{K}} &\equiv &\frac{16}{(1-u^{2})^{1/2}}>16,u\geqslant 0. 
\eqnum{3}
\end{eqnarray}%
Here, $u$ is the $t\rightarrow \infty $ velocity of the kink while $-u$ is
the $t\rightarrow \infty $ velocity of the antikink, and $v_{0},x_{0},\theta
\left( 0\leqslant \theta \leqslant 2\pi \right) $ are constants. Also, $%
\omega _{B}\equiv \cos \theta $ is the breather's internal frequency and $%
E,E_{K\stackrel{\_}{K}}$ are the breather and kink-antikink-pair energies,
respectively. Following Ref. [4], one assumes the{\it \ ansatz }that the
solution of the perturbed sine-Gordon equation has the same form as the
unperturbed breather (2), but now allowing $\theta $ and $v_{0}$ to be
temporal functions: 
\begin{equation}
v(t)=v_{0}(t)+\int^{t}\cos \theta \left( t^{\prime }\right) dt^{\prime }. 
\eqnum{4}
\end{equation}%
This approximation means that one is assuming a perturbation sufficiently
weak so that its principal effect on the system is to continuously alter the
frequency and the phase of the breather. Therefore, such an approximation is
valid when $\omega _{B}\ll 1$ since the breather energy is $E_{B}\equiv
16(1-\omega _{B}^{2})^{1/2}$. Next, one can deduce the equations of motion
for the breather collective modes $(u,v)$, where 
\begin{eqnarray}
u(t) &\equiv &U_{B}(x=x_{0},t)=4\tan ^{-1}A(t),  \nonumber \\
v(t) &\equiv &\left[ \frac{\partial U_{B}}{\partial t}\right] _{x=x_{0}}=%
\frac{4dA(t)/dt}{1+A\left( t\right) ^{2}},  \nonumber \\
A(t) &\equiv &\tan \theta \sin v\left( t\right) .  \eqnum{5}
\end{eqnarray}%
After the coordinate transformation $z\equiv \tan \left( u/4\right) ,%
\stackrel{.}{z}\equiv w,$ such equations of motion can be shown to be 
\begin{eqnarray}
\stackrel{.}{{\bf \chi }} &=&{\bf f}\left( {\bf \chi }\right) +\epsilon {\bf %
g}\left( {\bf \chi },\psi \right) ,  \nonumber \\
\stackrel{.}{\psi } &=&4K(m)/T,  \eqnum{6}
\end{eqnarray}%
with 
\begin{eqnarray}
{\bf \chi } &=&\left( 
\begin{array}{c}
z \\ 
w%
\end{array}%
\right)   \nonumber \\
{\bf f}\left( {\bf \chi }\right)  &=&\left( 
\begin{array}{c}
w \\ 
-z\left( 1-w^{2}\right) \left( 1+z^{2}\right) ^{-1}%
\end{array}%
\right) ,  \eqnum{7}
\end{eqnarray}%
and the perturbation 
\begin{equation}
\epsilon {\bf g}\left( {\bf \chi };t\right) =\left( 
\begin{array}{c}
0 \\ 
h(t)g_{p}(z,w)-\delta g_{\delta }(z,w)%
\end{array}%
\right) ,  \eqnum{8}
\end{equation}%
where 
\begin{eqnarray}
h(t) &=&\frac{\pi }{4}\gamma 
\mathop{\rm sn}%
\left[ 4K(m)t/T;m\right] ,  \nonumber \\
g_{p}(z,w) &=&\frac{\left[ \left( w^{2}+z^{2}\right) z^{2}-s\left\{
z^{4}+w^{2}\left[ z^{2}+\left( 1-w^{2}\right) \left( 1+z^{2}\right) \right]
\right\} \right] }{\left( 1+sw^{2}\right) \left( w^{2}+z^{2}\right) \left(
1+s\right) \left( 1+z^{2}\right) ^{1/2}}  \nonumber \\
&&+\frac{2w^{2}z^{2}s^{2}\left( 1+z^{2}\right) +w^{2}s^{3}\left(
1+2z^{2}\right) \left( 1+z^{2}\right) ^{2}}{\left( 1+sw^{2}\right) \left(
w^{2}+z^{2}\right) \left( 1+s\right) \left( 1+z^{2}\right) ^{1/2}}, 
\nonumber \\
g_{\delta }(z,w) &=&\frac{w\left[ \left( 2+z^{2}\right) \left(
w^{2}+z^{2}\right) z^{2}-s\left\{ z^{4}-w^{2}\left[ \left( 2+z^{2}\right)
z^{4}-\left( 1-w^{2}\right) \left( 1+z^{2}\right) \right] \right\} \right] }{%
\left( 1+sw^{2}\right) \left( w^{2}+z^{2}\right) \left( 1+z^{2}\right) } 
\nonumber \\
&&+\frac{w\left[ w^{2}s^{2}\left( 1+2z^{2}\right) \left( 1+z^{2}\right) %
\right] }{\left( 1+sw^{2}\right) \left( w^{2}+z^{2}\right) \left(
1+z^{2}\right) },  \nonumber \\
s &\equiv &\frac{\sinh ^{-1}(z)}{z\left( 1+z^{2}\right) ^{1/2}}.  \eqnum{9}
\end{eqnarray}%
Noting that the separatrix in the phase-plane $u-v$ is given by 
\begin{eqnarray}
u_{0,\pm }(t) &=&\pm 4\tan ^{-1}(t),  \nonumber \\
v_{0,\pm }(t) &=&\pm 4/(1+t^{2}),  \eqnum{10}
\end{eqnarray}%
one finds that on the separatrix in the phase-plane $z-w$ 
\begin{eqnarray}
z_{0,\pm }(t) &=&\pm t,  \eqnum{11} \\
w_{0,\pm }(t) &=&\pm 1,  \nonumber
\end{eqnarray}%
where the positive (negative) sign refers to the upper (lower) homoclinic
orbit in the phase-plane $u-v$, which is obtained for the energy $E=16$ of
the unperturbed energy functional $E=2\tan ^{2}(u/4)+\left( \stackrel{.}{u}%
/4\right) ^{2}\sec ^{2}(u/4).$

\section{\protect\bigskip Chaotic threshold function}

Now one can estimate the threshold of chaotic instabilities for the
effective system (6) using Melnikov's method. Thus, the corresponding
Melnikov function is 
\begin{equation}
M^{\pm }(t_{0})=\mp D+C\sum_{n=0}^{\infty }a_{n}(m)p_{n}(T)\sin \left[ \frac{%
\left( 2n+1\right) 2\pi t_{0}}{T}\right] ,  \eqnum{12}
\end{equation}%
where

$\bigskip $%
\begin{eqnarray}
D &\equiv &\delta \int_{-\infty }^{\infty }\frac{\left( 1-\tanh ^{2}x\right) %
\left[ \left( 2-\tanh ^{2}x\right) \tanh ^{2}x+r+r^{2}\left( 1-\tanh
^{4}x\right) \right] }{1+r}dx\simeq 2.03727\delta ,  \nonumber \\
p\left( 2\pi /T\right)  &\equiv &\int_{-\infty }^{\infty }\frac{\cos \left( 
\frac{2\pi }{T}\sinh x\right) \left[ \left( 1-r\right) \left( 1-\tanh
^{2}x\right) \right] 
\mathop{\rm sech}%
x+2r^{2}\tanh ^{2}x+r^{3}\left( 1+\tanh ^{2}x\right) }{\left( 1+r\right) ^{2}%
}dx,  \nonumber \\
C &\equiv &\pi ^{2}\gamma /(4\sqrt{m}K(m)),  \nonumber \\
a_{n}(m) &\equiv &%
\mathop{\rm csch}%
\left[ \left( 2n+1\right) \pi K(1-m)/(2K(m))\right] ,  \nonumber \\
p_{n}(T) &\equiv &p\left[ \left( 2n+1\right) 2\pi /T\right] ,  \eqnum{13}
\end{eqnarray}%
and where $r\equiv 2x%
\mathop{\rm csch}%
\left( 2x\right) $. From Eq. (12) one sees that a homoclinic bifurcation is
guaranteed if 
\begin{equation}
\frac{\delta }{\gamma }<U\left( m,T\right) ,  \eqnum{14}
\end{equation}%
where the chaotic threshold function is 
\begin{equation}
U\left( m,T\right) \equiv \frac{\pi ^{2}}{4\alpha \sqrt{m}K(m)}%
\sum_{n=0}^{\infty }\left( -1\right) ^{n}a_{n}\left( m\right) p_{n}(T). 
\eqnum{15}
\end{equation}%
The series in Eq. (15) satisfies the conditions of the Leibnitz theorem on
alternate series, and then one can deduce a slightly more restrictive
condition for the onset of the aforementioned spatio-temporal chaotic
instabilities, in order to facilitate the comparison with the previously
studied case [4] of a harmonic force $\left( m=0\right) $: 
\begin{equation}
\frac{\delta }{\gamma }<U\left( m=0,T\right) R(m),  \eqnum{16}
\end{equation}%
where $U(m=0,T)=\pi p(2\pi /T)/(4\alpha ),R(m)\equiv \pi \left( \sqrt{m}%
K(m)\right) ^{-1}%
\mathop{\rm csch}%
\left[ \pi K(1-m)/(2K(m))\right] $. It is worth noting that the condition
(16) is a reliable approximation due to the coefficients $a_{n}(m)\simeq 0$, 
$\forall n\geqslant 1$ and $\forall m\in \left[ 0,m^{\ast }\right] $ with $%
m^{\ast }$ being very close to 1. A plot of the shape function $R(m)$ is
shown in Fig. 1, where one sees that $R(m=0)=1$, as expected. The rapid
growth of $R(m)$ as $m\rightarrow 1$ means that the chaotic threshold
exhibits hypersensitivity to changes in the excitation shape as this
approximates a square wave. This can be understood in terms of the
increasing accumulation of {\it effective} harmonics in the Fourier series
of the elliptic force as $m$ $\rightarrow $1. Note that this number of
effective harmonics is not a linear function of $m$ because of the
dependence of the elliptic force on $K(m)$.

\section{Numerical simulations}

For the purpose of the comparison with the above theoretical predictions,
the parameters $\delta =0.2,T=2\pi /0.6,$and $L=24$ were fixed, and an sG
breather with frequency $\omega _{B}=\sqrt{2}/2$ and $x_{0}=v_{0}=0$ (cf.
Eq. (2)) was chosen as initial data. The period $T=2\pi /0.6$ corresponds to
a frequency $\omega =0.6$ (for a sinusoidal force, $m=0$) for which the
chaotic attractors are dominated by breather to kink-antikink transitions
[11,12]. Equation (1) was integrated under periodic boundary conditions $%
\Phi \left( x=-L/2,t\right) =\Phi \left( x=L/2,t\right) $ for all $t$.
Figure 2 shows an illustrative example of the comparison between the
theoretically obtained chaotic threshold amplitude $\gamma _{th}$ (solid
line, cf. Eq. (16)) and the corresponding numerically obtained amplitude
(dots) as the shape parameter $m$ varies. One sees that the amplitudes
obtained numerically are slightly larger than the respective theoretically
predicted ones, as expected, although an overall good agreement is found.
Similar good agreement was found for other sets of parameters in the
large-period regime ($T\gg 2\pi $), provided that the damping and driving
force were sufficiently weak. An illustrative sequence of the order-chaos
route is shown in Fig. 3 where the initial state ($m=0$) occurs after the
system has undergone a spatial period doubling (for $m=0$ and $0.6\lesssim
\gamma \lesssim 0.9$, cf. Ref. [12]) but remains simply periodic in time.
This two-breather state has thus a spatial period $\frac{1}{2}$ and a
temporal period 1, and exhibits mirror symmetry with respect to the origin.
The spatial symmetry is broken as the forcing shape deviates from
sinusoidal. Figure3 illustrates this phenomenon with the values $m=0.8$ and $%
m=0.82$. This spatial symmetry breaking of the two-breather state appears to
be an unavoidable precursor of the onset of spatio-temporal chaos via the
kink-antikink collision mechanism, as can be appreciated in Fig. 3 for the
shape parameter $m=0.898\simeq m_{th}$, which corresponds to the
theoretically predicted chaotic threshold (cf. Eq. (16)). For higher values
of the shape parameter, i.e., when the driving shape is ever closer to a
square wave, one finds the persistence of the chaotic dynamics (cf. Figs. 3, 
$m=0.99$) predicted by the Melnikov-method-based analysis.

\section{\protect\bigskip Conclusions}

In sum, it has been shown through the example of an elliptic ac driving that
the competition between sine-Gordon-breather and kink-antikink-pair solitons
strongly depends on the shape of the spatially uniform ac driving. In
particular, it was theoretically predicted and numerically confirmed that a
square wave is more effective than a sinusoidal function to yield
spatio-temporal chaos associated with the competition between
sine-Gordon-breathers and kink-antikink-pairs. Potential application of the
present results to reshaping-induced control (suppression and enhancement)
of chaotic dynamics [1,14,15] range from DNA molecules [16] to long
Josephson junctions and quasi-one-dimensional ferromagnets [17]. An
interesting problem under current investigation is whether the above
predictions on the chaotic threshold remain reliable in the so-called (for $%
m=0$) nonlinear (cubic) Schr\"{o}dinger regime (i.e., $T\lesssim 2\pi $),
where the chaotic dynamics is dominated by breather-breather transitions
instead of the breather to kink-antikink transitions of the regime studied
here ($T\gg 2\pi $), since both mechanisms involve homoclinic crossings [13].

\begin{acknowledgement}
The author acknowledges financial support from the Spanish MCyT and the
European Regional Development Fund (FEDER) program through project
FIS2004-02475.
\end{acknowledgement}

\bigskip 

\subsubsection{Figure Captions}

\bigskip 

Figure 1. Shape function $R(m)$ (see the text) versus the shape parameter $m$%
.

\bigskip 

Figure 2. Theoretical (solid line, cf. Eq. (16)) and numerically obtained
(dots) chaotic threshold amplitude, $\gamma _{th}$, versus shape parameter, $%
m$, for $\delta =0.2$ and $T=2\pi /0.6$.

\bigskip 

Figure 3. Space-time evolution of $\Phi \left( x,t\right) $ for the sG
system through four driving periods for $\delta =0.2,\gamma =0.79,T=2\pi /0.6
$, and shape parameters $m=0$, $0.5$, $0.8$, $0.82$, $0.898\simeq m_{th}$,
and $0.99$.

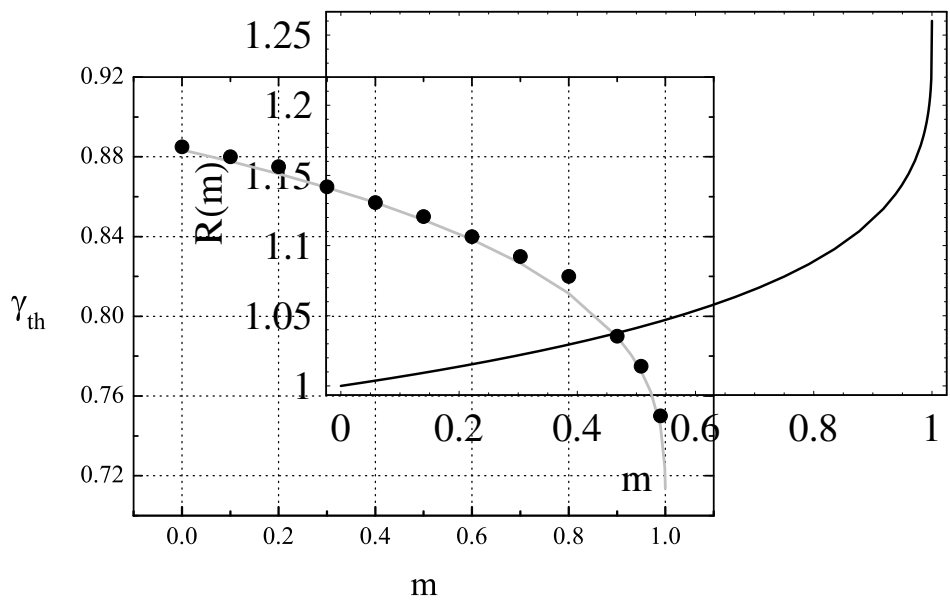
\end{document}